\PassOptionsToPackage{dvipsnames}{xcolor}

\documentclass{aastex6}
\usepackage{mathtools}
\usepackage{amsmath}
\usepackage{float}
\usepackage{longtable}
\usepackage{soul}
\usepackage{multirow}
\usepackage{savesym}
\savesymbol{tablenum}
\usepackage{siunitx}
\usepackage{tabularx}
\usepackage{makecell}
\usepackage[toc,page]{appendix}
\let\tablenum\relax
\usepackage{siunitx}
\usepackage{natbib}
\usepackage{hyperref}

\fullcollaborationName{The Friends of AASTeX Collaboration}

\begin{document}

\title{Imaging the Inner Astronomical Unit of Herbig Be Star HD 190073}

\author{Nour Ibrahim\altaffilmark{1}, John D. Monnier\altaffilmark{1}, Stefan Kraus\altaffilmark{2}, Jean-Baptiste Le Bouquin\altaffilmark{3}, Narsireddy Anugu\altaffilmark{4},Fabien Baron\altaffilmark{3,6}, Theo Ten Brummelaar\altaffilmark{4}, Claire L. Davies\altaffilmark{2}, Jacob Ennis\altaffilmark{1}, Tyler Gardner\altaffilmark{1}, Aaron Labdon\altaffilmark{5}, Cyprien Lanthermann\altaffilmark{4}, Antoine Mérand\altaffilmark{7}, Evan Rich\altaffilmark{1}, Gail H. Schaefer\altaffilmark{4}, Benjamin R. Setterholm\altaffilmark{1}}

\altaffiltext{1}{Astronomy Department, University of Michigan, Ann Arbor, MI 48109, USA}
\altaffiltext{2}{Astrophysics Group, Department of Physics \& Astronomy, University of Exeter, Stocker Road, Exeter, EX4 4QL, UK}
\altaffiltext{3}{Institut de Planetologie et d'Astrophysique de Grenoble, Grenoble 38058, France}
\altaffiltext{4}{The CHARA Array of Georgia State University, Mount Wilson Observatory, Mount Wilson, CA 91203, USA}
\altaffiltext{5}{European Southern Observatory, Casilla 19001, Santiago 19, Chile}
\altaffiltext{6}{Department of Physics and Astronomy, Georgia State University, 25 Park Place NE 605, Atlanta, GA 30303-2911, USA}
\altaffiltext{7}{European Organisation for Astronomical Research in the Southern Hemisphere (ESO), Karl-Schwarzschild-Str. 2, 85748 Garching
bei München, Germany}
\begin{abstract}

Inner regions of protoplanetary disks host many complex physical processes such as star-disk interactions, magnetic fields, planet formation, and the migration of new planets. To directly study this region requires milli-arcsecond angular resolution, beyond the diffraction limit of the world’s largest optical telescopes and even too small for the mm-wave interferometer ALMA. However, we can use infrared interferometers to image the inner astronomical unit. Here, we present new results from the CHARA and VLTI arrays for the young and luminous Herbig Be star HD 190073. We detect a sub-AU cavity surrounded by a ring-like structure that we interpret as the dust destruction front. We model the shape with 6 radial profiles, 3 symmetric and 3 asymmetric, and present a model-free image reconstruction. All the models are consistent with a near face-on disk with inclination $\lesssim 20^\circ$, and we measure an average ring radius of 1.4$\pm0.2$ mas (1.14 AU). Around 48\% of the total flux comes from the disk with ~15\% of that emission appearing to emerge from inside the inner rim. The cause of emission is still unclear, perhaps due to different dust grain compositions or gas emission. The skewed models and the imaging point to an off-center star, possibly due to binarity. Our image shows a sub-AU structure, which seems to move between the two epochs inconsistently with Keplerian motion and we discuss possible explanations for this apparent change. 

\end{abstract}

\section{Introduction} 
\label{sec:intro}

As the catalogs of exoplanets grow, we increasingly see evidence of otherworldly environments that we do not observe in our local solar system \citep[see][]{cacc2022,Rein2012}. Observing the early formation stages of planetary systems is key to understanding and eventually predicting the worlds we continually find. The planet formation process is still not well understood, especially around massive stars. Herbig Ae/Be stars are a class of intermediate to high-mass (2-10  M$_{\odot}$) pre-main sequence stars of spectral type A or earlier that exhibit excess near-infrared (NIR) emission, which is associated with circumstellar disks. Disks around the more massive B stars are harder to study because they disappear on shorter timescales, around a few $10^5$ yr, likely due to photoevaporation caused by UV radiation from the star. Herbig Be stars, in particular, are often observed with a factor of 5-10 lower disk masses than Herbig Ae and T Tauri stars, or in some cases, are not detected at all \citep{Alonso2009}. The inner regions of these disks are particularly interesting because they host many complex processes that go into planet formation, yet they are not well studied. Imaging the sub-astronomical unit (sub-AU) requires milli-arcsecond angular resolution, which is not achievable using the world's largest optical telescopes and is even too small for the mm-wave interferometer ALMA. However, infrared long-baseline interferometers can probe the disks at the sub-AU scale and give us a better look at the inner regions of the disks. 

Modeling has been a key tool for studying the structure of circumstellar accretion disks. Early models by \citet{Hillenbrand1992}, which assumed a flat and optically thick disk extending a few stellar radii close to the star, were able to reproduce photometric near-infrared excess emission measurements. However, this theoretical picture was not confirmed by the interferometric observations carried out by \citet{Millan1999}, which found that the disk sizes had to be much larger than what was predicted by the thin and optically thick disk models. In 2001, \citet{natta2001} and \citet{Dullemond2001} proposed a new model for the inner disk region, suggesting an optically thin cavity around the star and a  ``puffed-up" inner rim wall that is truncated at a sub-AU radius where the rim temperature is equal to the dust sublimation temperature. Further observations by \citet{2001millan} and \citet{Tuthill2001} showed that this model where the majority of the NIR excess originates from the dust sublimation radius, explained the large photometric NIR bump. \citet{Monnier2002} validated the ``puffed-up” inner rim with the introduction of the size-luminosity diagram where they also showed that the dust sublimation temperature was between 1500-2000 K for their sample. However, more recent sub-milliarcsec interferometry observations of AB Aurigae and MWC 275 (HD 163296) carried out by \citet{Tannirkulam2008} showed that models in which the dust evaporation rim solely produces the NIR excess, fail to explain the data. They found that a significant amount of the inner emission emerges from within the dust sublimation radius which likely does not have a sharp edge. \citet{Lazareff2017}’s PIONIER survey confirmed this general picture in addition to adding 27 new objects to the size-luminosity diagram by \citet{Monnier2002} and found some variations between sources.

Most of what we know about Herbigs comes from observations of the more common and close-by Herbig Ae stars. Due to the rareness of massive stars and the very short disk lifetimes, only a handful of Herbig Be stars with detected disks are close enough for us to study, one of which is the B9 star HD 190073, also known as V1295 Aquila (see stellar parameters in \autoref{table:stellar params}) \citep{Rich2022}. HD 190073 is a pre-main sequence star with a spectral type of B9 and a temperature of approximately 9750K. It has a luminosity of $\sim760$ L$_\odot$ and a model-derived mass of 6.0 $\pm 0.2$ M$_\odot$. Based on its mass, this star is expected to rapidly contract onto the main sequence with a spectral type of B5 and a temperature of 15500 K in $\sim 10^{5} $ years \citep{2000Cox}. While HD 190073 shares some similarities with other well-studied pre-main sequence stars such as Herbigs AB Aur and MWC 275, it has more than twice the mass and will likely have a significantly different experience in terms of its disk evolution. It also has a detected magnetic field which is uncommon among Herbigs \citep{Alecian2013}. The magnetic field is much weaker than those of T Tauri stars, but it could be measured due to the star's narrow spectral lines. The low  $v sin i ~ 8.6$ km/s is partly explained by a face-on geometry \citep{Catala2007}. Unfortunately, ALMA has not observed it yet so we cannot confirm through imaging that the disk is face-on. 2D models from the PIONIER/VLTI survey found that HD 190073 is nearly face-on and symmetric, but they did not have enough resolution to see any inner emission \citep{Lazareff2017}. The results from this PIONIER study revealed that the inclination of the HD 190073 disk is $< 20^{\circ}$, and therefore it is plausible that the  slow $v\sin{i}$ is due to the low inclination.

\citet{Setterholm2018} recently presented the first CHARA results on HD 190073 with 3 times higher resolution. They used broadband data taken over many years and combined them as a single epoch. The resulting visibility curves lacked the expected bounce at larger baselines associated with a ring-like structure. The HD 190073 models showed a face-on disk that favored a Gaussian shape rather than a thin ring. This conclusion was similar to results from \citet{Tannirkulam2008} for the older, less massive, and less luminous Herbig Ae stars AB Aur and MWC 275.

Here in this work, we will revisit this analysis using new ``snapshot'' observations taken over a much shorter period of three months and with a much denser (u,v) coverage.  We obtained the new H-band interferometry using the updated Michigan InfraRed Combiner - eXeter (MIRC-X)  \citep{anugu2020} instrument at the Center for High Angular Resolution Astronomy (CHARA) Array \citep{Brummelaar2005} and Precision Integrated-Optics Near-infrared Imaging ExpeRiment (PIONIER) \citep{LeBouquin2011} at the Very Large Telescope Interferometer (VLTI). With improved angular resolution and increased (u,v) coverage, we present new models with six different radial profiles along with a model-free image reconstruction of the sub-AU ring. The enhanced data quality allows us to model finer details in the disk structure and interpret the inner region dynamics more accurately.

We start by describing our observation routines to collect the CHARA and VLTI data, as well as discussing our data reduction techniques in section \autoref{Sec:Obser}. Next, we explain our methods for combining the data from the two instruments into two epochs, and we show a basic presentation of the (u,v) coverage, visibility, and closure phase measurements for each epoch in section \autoref{sec:DATA}. We present our symmetric and asymmetric models in \autoref{sec:Modeling} and describe them extensively. Then in section \autoref{sec:imaging} we present model-free images and compare them to the models. Finally, we discuss in section \autoref{sec:disc} what the results of the modeling and imaging reveal about the inner disk of HD 190073.

\newpage
\begin{table}[h]
\centering
\begin{tabular}{llll}
\hline
Property      &  &  & HD 190073  \\     

\hline
$\alpha$ (J2000) &  &  & $20^h03^m02^s.51$   \\           
$\delta$ (J2000) &  &  & +5\textdegree44'16''.66  \\   
Spectral Type &  &  & B9               \\
T$_{eff}$          &  &  & 9750 $\pm 125$ K \\
R$_*$         &  &  & 9.68 $\pm 0.44$ R$_\odot $  \\
Age           &  &  & 0.30 $\pm 0.02$ Myr\\
Distance      &  &  &  824.16 $\pm 21.83 $ pc \\
Log(L$_*$) L$_\odot$    &   &  & 2.88 $\pm$ 0.03 \\  
Mass          &  &  & 6.0 $\pm 0.2$ M$_\odot$ \\
Vmag          &   &  & 7.79 $\pm 0.06$ \\
Hmag          &   &  & 6.61 $\pm 0.07$ \\

\end{tabular}
\caption{\label{tab:table-name}HD 190073 Stellar properties from \citet{guzman2020} }
\label{table:stellar params}
\end{table}

\section{Observations and Data Reduction} 
\label{Sec:Obser}
This paper relies on new infrared interferometry from two different facilities, CHARA and VLTI. We coordinated observations between Mt.Wilson in California and Paranal Mountain in Chile using similar wavelengths for one month to get better-quality data. It was also important to get data around the same time due to time variability because of the short rotational period of the sub-AU region of the disk $\sim 0.8$ years, which we calculated from previous estimations of the inner radius. Interferometric time variability has been reported for other young stellar objects (YSOs), and it was shown to be present in $>10\%$ of the accretion disk sample studied by \citet{kobus2020}. The goal of the experiment is to make the best quality images and models by combining the baselines and (u,v) coverage of the two facilities. We were granted observing time over a three-month period, so we expect to be able to resolve some sub-AU disk motion if present. We have six observations using the CHARA array's MIRC-X instrument, and seven observations using the VLTI array's PIONIER instrument. In this section, we describe the observations at the two facilities.

\subsection{CHARA}

HD 190073 was recorded using the MIRC-X instrument in the summer of 2019 as listed in \autoref{table:chara}. MIRC-X is an infrared six-telescope beam combiner at the CHARA Array. The CHARA interferometer is located at Mt. Wilson in California, USA, with an array of six 1-meter telescopes disposed in a Y shape, optimizing the imaging capability. The telescopes combine to have baselines of up to 331 meters. The maximum angular resolution is $\frac{\lambda}{2B}$ which corresponds to 0.51 mas in H-band ($\lambda = 1.64\,\mu m$). MIRC-X uses single-mode fibers to coherently combine the light from six telescopes simultaneously with an image-plane combination scheme and typically delivers a visibility precision better than 5$\%$, and closure phase precision better than 1$^{\circ}$. The data were collected using the spectral resolution $R=50$ mode using a prism dispersive element giving 8 spectral channels spread over $\Delta\lambda = 0.27\mu m$. We reduced the data using the MIRC-X reduction pipeline \footnote{\url{https://gitlab.chara.gsu.edu/lebouquj/mircx_pipeline}} (version 1.3.3) and calibrated the data using a custom IDL routine \footnote{Contact \href{mailto:monnier@umich.edu}{monnier@umich.edu} for IDL routine}. The MIRC-X data reduction pipeline produces science-ready visibilities and closure phases written in OIFITS format (\citealp{Duvert2017,Pauls2005}). The raw visibility is calibrated using stars with known diameters. Typical observations execute a standard calibrator-science-calibrator (CAL-SCI-CAL) cycle. Calibrators are usually chosen, using the SearchCal software, to be close to the science target, both in terms of sky position and magnitude, and have a smaller angular diameter so that their visibility on a given baseline is less dependent on the diameter \citep{chelli2016}.

\begin{table}[H]
\centering
\caption{CHARA/MIRC-X Observations of HD190073}


\begin{tabular}{llcccm{20mm}m{20mm}c}


\multicolumn{1}{c}{UT Date} & \multicolumn{1}{c}{Configuration} & Band & no. $\mathcal{V}^2$ ($\times$ 8) & no. CP ($\times$ 8)    & \multicolumn{1}{c}{Calibrator(s)}  & \multicolumn{1}{c}{Diameter (mas)}                                 &    Epoch\\     
    \hline
    2019-05-09 & E1-W2-W1-S1-E2 & H & 9 & 7& HD 190753& 0.323$\pm$0.008 & A* \\ 
    \hline
    2019-06-05   & E1-W2-W1-S2-S1-E2 & H & 67
    & 77 
    & \begin{tabular}[c]{@{}l@{}}HD 190753\\ HD 191656\end{tabular} & \begin{tabular}[c]{@{}l@{}}0.323$\pm$0.008\\0.430$\pm$0.01 \end{tabular} &A \\ 
    \hline
    2019-06-09 & E1-W2-W1-E2 & H    & 26
    & 24 
    & HD 197551&0.644$\pm$0.02  &A \\ 
    \hline
    2019-07-11 & E1-W2-W1-S2-S1-E2 & H    & 66 
    & 74 
    & HD 190753           & 0.323$\pm$0.008      &B  \\
    \hline
   2019-07-12 & E1-W2-W1-S2-E2 & H    & 25
   & 23
   & HD 190753           & 0.323$\pm$0.008      &B  \\
    \hline
    2019-08-27 & E1-W2-W1-S2-S1-E2 & H    & 36 & 46  & \begin{tabular}[c]{@{}l@{}}HD 191656\\ HD 190753\end{tabular}   & \begin{tabular}[c]{@{}l@{}}0.430$\pm$0.01\\0.323$\pm$0.008\end{tabular}    &B*\\
    \hline
\end{tabular}
\label{table:chara}
 \\[10 pt] 
  *For imaging, this epoch was not included; see \autoref{sec:imaging}

\end{table}

\subsection{VLTI}
 HD 190073 was recorded with the PIONIER instrument on multiple occasions throughout 2019, as listed in  \autoref{table:vlti}. PIONIER is a four-telescope beam combiner operating in the H-band ($\lambda = 1.64\,\mu m$) at the VLTI. Data were obtained using the four auxiliary telescopes (ATs) in multiple baseline configurations giving baselines ranging from 11.3\,m to 140.0\,m, allowing for a maximum spatial resolution ($\lambda/B$) of 1.12\, mas. Additionally, a prism dispersive element was used giving 6 spectral channels spread over $\Delta\lambda = 0.30\,\mu m$. Observations were taken in concatenations of calibrator (CAL) and science (SCI) targets in blocks of either CAL-SCI-CAL or CAL-SCI-CAL-SCI-CAL, to allow for effective monitoring of the transfer function and precise calibration. The data were reduced and calibrated using the standard \textsc{PNDRS} package v3.52 \citep{LeBouquin2011}.

\begin{table}[H]
\centering
\caption{VLTI/PIONIER Observations of HD190073}
\begin{tabular}{llcccm{20mm}m{20mm}l}
\multicolumn{1}{c}{UT Date} & \multicolumn{1}{c}{Configuration} & Band & no. $\mathcal{V}^2$ ($\times$ 6)  & no. CP ($\times$ 6)     & \multicolumn{1}{l}{Calibrator(s)}                          & \multicolumn{1}{l}{Diameter (mas)}                                 &    Epoch\\ 

\hline
2019-06-09               & D0-G2-J3-K0  (Medium)                    & H    & 12  & 8             & \begin{tabular}[c]{@{}l@{}}HD 188385\\HD 189509\end{tabular} & \begin{tabular}[c]{@{}l@{}}0.221$\pm$0.01\\0.247$\pm$0.03\end{tabular} &A \\ 
\hline
2019-07-10               & D0-G2-J3-K0  (Medium)                 & H    & 24 & 16             & \begin{tabular}[c]{@{}l@{}}HD 190753\\HD 191840\end{tabular} & \begin{tabular}[c]{@{}l@{}}0.323$\pm$0.02\\0.288$\pm$0.01\end{tabular} &A \\ 
\hline
2019-07-20               & A0-G1-J2-J3  (Large)                     & H    & 30  & 20            & \begin{tabular}[c]{@{}l@{}}HD 190753\\HD 191840\end{tabular} & \begin{tabular}[c]{@{}l@{}}0.323$\pm$0.02\\0.288$\pm$0.01\end{tabular} &A\& B \\ 
\hline
2019-07-30               & A0-B2-C1-D0  (small)                     & H    & 24 & 16             & \begin{tabular}[c]{@{}l@{}}HD 190753\\HD 191840\end{tabular} & \begin{tabular}[c]{@{}l@{}}0.323$\pm$0.02\\0.288$\pm$0.01\end{tabular}  &A\\ 
\hline
2019-08-04               & A0-B2-C1-D0  (small)                     & H    & 12 & 8            & \begin{tabular}[c]{@{}l@{}}HD 190753\\HD 191840\end{tabular} &  \begin{tabular}[c]{@{}l@{}}0.221$\pm$0.01\\0.247$\pm$0.03\end{tabular}  &B\\ 
\hline
2019-08-05               & D0-G2-J3-K0  (Medium)                    & H    & 6 & 4             & \begin{tabular}[c]{@{}l@{}}HD 188385\\HD 189509\end{tabular} & \begin{tabular}[c]{@{}l@{}}0.221$\pm$0.01\\0.247$\pm$0.03\end{tabular}  &B\\ 
\hline
2019-08-06               & D0-G2-J3-K0  (Medium)                    & H    & 6 & 4             & \begin{tabular}[c]{@{}l@{}}HD 188385\\HD 191840\end{tabular} & \begin{tabular}[c]{@{}l@{}}0.221$\pm$0.01\\0.288$\pm$0.01\end{tabular}  &B\\
\hline
\end{tabular}
\label{table:vlti}
\end{table}

\section{Basic Data Presentation}
\label{sec:DATA}
\subsection{Combining VLTI and CHARA Data}
\label{subsec:combine}
Using the angular resolution of the instruments and the Keplerian velocity of the inner disk, we calculated the timescale on which each instrument is able to resolve a change in the disk (eg. rotation).
CHARA has a maximum angular resolution of $\frac{\lambda}{2B}$ = 0.51 mas for $\lambda =1.65$ $\mu m$ and $B = 331$ m. We estimated the Keplerian velocity of the dust rim using R $\sim 1.64$ AU from \citet{Setterholm2018} and the star's mass $M = 6$ M$_{\odot}$ which gave us a rotational period of T = 304.5 days or 0.83 years. That means that it takes 13 days for us to be able to resolve the motion of one resolution element using MIRC-X. Doing the same calculation for the large VLTI baseline, we found that it takes 28 days to resolve a change in one resolution element using PIONIER. With that information, we aimed to split the data into two epochs, such that the amount of time in which we expect each epoch to change is minimized. We grouped the earlier three CHARA nights to make epoch A and the later 3 nights to make epoch B. When combining VLTI nights, we had to consider the baseline configuration as well as the dates. The different configurations (small, medium, and large) correspond to different (u,v) coverage. The small configuration contributes to the short baselines which show us the large-scale components of the disk, while the large configuration, which contributes the long baselines, allows us to constrain the small-scale components of the inner region. To get the most coverage, we included the first 4 nights (2019-06-09, 2019-07-10, 2019-07-20, 2019-07-30) into epoch A. Even though the fourth night, 2019-07-30, is more than 28 days away from the first night, 2019-06-09, we included it in epoch A because it was the closest night with a small baseline configuration. The shorter the baseline, the longer the timescale for resolving a significant change. That timescale for the small configuration is $>$ 1 month, so adding 2019-07-30 to epoch A is justified. The last 3 nights (2019-08-04, 2019-08-05, 2019-08-06) went into epoch B, as well as 2019-07-20 which is shared between the two epochs as it is the only large baseline configuration we have. The last column of \autoref{table:chara} and \autoref{table:vlti} indicates the epoch to which each night corresponds. We show the combined full (u,v) coverage, squared visibilities ($\mathcal{V}^2$), and closure phases (CP) from both instruments for each epoch in \autoref{fig:uv}.

Based on early studies of MIRC \citep{Monnier2012}, we adopted the same minimum systematic errors. There are two types of errors associated with ($\mathcal{V}^2$) measurements. There are multiplicative errors due to the effects of seeing, and there are additive errors correcting for biases due to the limitations of the pipeline. The multiplicative error is $6 \%$ while the the additive error is  $2\times10^{-4}$. These are used to calculate a new minimum error. If the systematic error is larger than the statistical one, then we adopt the larger one. The same thing is done for triple amplitudes but the associated errors are $10 \%$ for multiplicative and $1\times10^{-5}$ for additive. Closure phases have an error floor of $0.1^{\circ}$. 

In section \autoref{sec:DATA}, we discuss our method of reducing the data and combining the CHARA and VLTI nights into two epochs.

\subsection{Wavelength Interpolation}
\label{sec:wlsmooth}

Since the wavelength channels vary between MIRC-X and PIONIER, and in fact, even between nights on the same instrument, we fit each the square visibility and closure phase measurement as a function of wavelength to a quadratic with a linear regression and re-sampled our data at standard wavelengths in the range 1.5-1.7 $\mu m$ at $50$ nm increments. This wavelength ``smoothing" produces evenly spaced visibility measurements in 5 wavelength  channels which we then use for modeling and imaging in the following sections. The amplitude of the error estimates was also fit to a quadratic and re-sampled at the chosen wavelengths. We then applied the same minimum error calibration as described in \autoref{subsec:combine}

\subsection{Squared Visibilities}
The combination of CHARA and VLTI allows us to sample the (u,v) coverage well. Just looking at the non-wavelength-smoothed visibility curve in \autoref{fig:uv}, we can visually detect a bounce that was not as clear in previous studies \citep{Setterholm2018}. Due to the short time between observation nights, we are able to avoid some smearing effects that the rotation of the disk can cause on longer time scales. As mentioned in \autoref{sec:wlsmooth} above, due to the mismatch in wavelengths between nights and instruments, we will be using the wavelength-smoothed data for modeling in the next section. 
An initial qualitative look at the visibility curves can already point to features that we can predict. The short baseline measurements from PIONIER appear to be slightly less than unity at 0 baseline, which could indicate the presence of an over-resolved large-scale halo structure. As the baseline increases, the visibility drops off rapidly at the short baselines. The drop-off in visibility tells us about the disk diameter. Towards longer baselines from MIRC-X, we see a bounce in the visibility. A bounce in the intermediate baselines indicates the existence of a cavity in the disk, and the severity of the bounce characterizes the sharpness of the rim. A shallow bounce points to a fuzzy rim, while a steep bounce points to a sharper transition \citep{Dullemond2010}. Fits by \citet{Lazareff2017} of HD 190073 seemed to favor a ring structure like we expect to see from the visibility curves, while fits by \citet{Setterholm2018} were not able to recover a cavity and concluded a Gaussian shape instead.
 
\subsection{Closure Phases}
\begin{figure}[h]

\centering
\includegraphics[trim={0cm 5cm 0 6.55cm},clip,width=0.94\textwidth,scale=1 ]{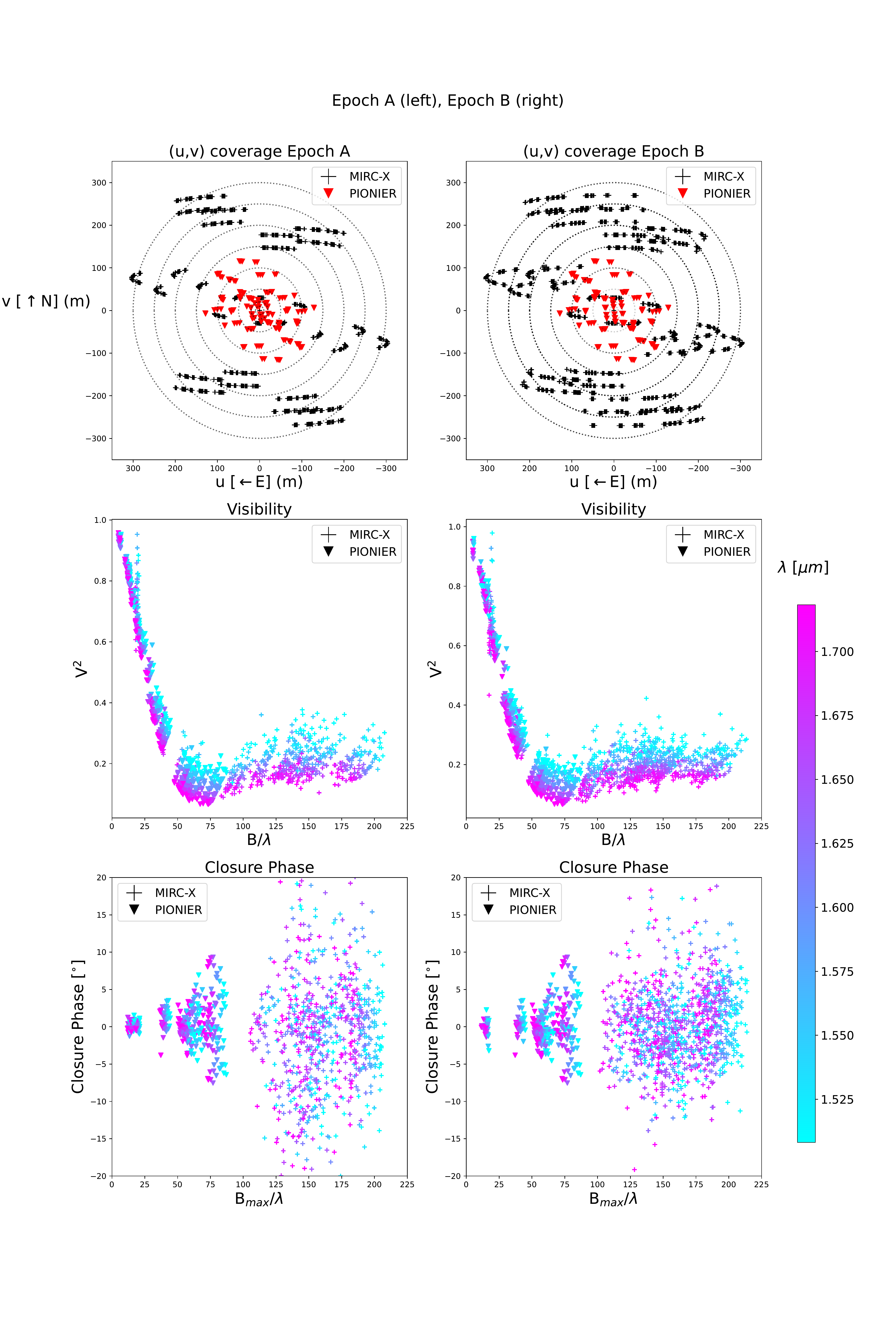}
\caption{Top row: The (u,v) coverage of the combined data for epochs A (left) and B (right). Data from MIRC-X are shown in black, and PIONIER are shown in red. Middle row: Unsmoothed combined squared visibility measurements from both instruments. The colors correspond to the H-band wavelengths with the darkest being the longest wavelength. Bottom row: Closure phase measurements from both instruments }
\label{fig:uv}
\end{figure}

Closure phase measurements from both instruments and for both epochs can be seen in the bottom two panels of \autoref{fig:uv}. The majority of the closure phases are relatively small $\lesssim 10^{\circ}$ which tell us that the disk is not far from symmetric. On the shorter baselines, the closure phase measurements are closest to zero, which tells us that the larger scale brightness distribution is nearly symmetric. However, since the closure phases at longer baselines are non-zero and non-180$^{\circ}$ we expect the disk to have skewness on the finer spatial scale, albeit small.

Similar to the visibility plots, we are not showing the error bars on the measurements. That was done mainly to show a simplistic visualization of what the data looks like. Typical errors are $\lesssim 5^\circ$.

\newpage
\section{Modeling}
\label{sec:Modeling}
We split our simple geometric models into two categories in this section. First, we assume point symmetry and fit the squared visibilities and not the closure phases since the disk is nearly face-on and we have small closure phase measurements. The symmetrical models will allow  us to compare our modeling to previous work, as well as constrain some physical parameters. Second, we allow the models to be asymmetric by fitting both the squared visibilities and closure phases and adding more fitting parameters to test which model best fits the closure phases. All modeling was done using PMOIRED \footnote{\url{https://github.com/amerand/PMOIRED}}\citep{2022merand}. In order to combine data from all wavelengths, we modeled the star as a point source with a fixed power spectrum proportional to $\lambda^{-4}$. The second component that went into the model is the disk with a power spectrum $\beta_{disk} \lambda ^{\alpha_{disk}}$ and we let $\alpha_{disk}$ and $\beta_{disk}$ be free parameters. We added a resolved halo component to the models to account for any large-scale scattered light which helps us constrain the flux contributions better. This halo component had a similar power spectrum to the disk $\beta_{halo} \lambda ^{\alpha_{halo}}$ and we let $\alpha_{halo}$ and $\beta_{halo}$ be free parameters. The $\alpha_{disk}$ and $\alpha_{halo}$ are what we are going to refer to as spectral slopes in the next section.

\subsection{Symmetric Modeling}
\begin{figure}[H]
\centering
\includegraphics[width=0.86\textwidth]{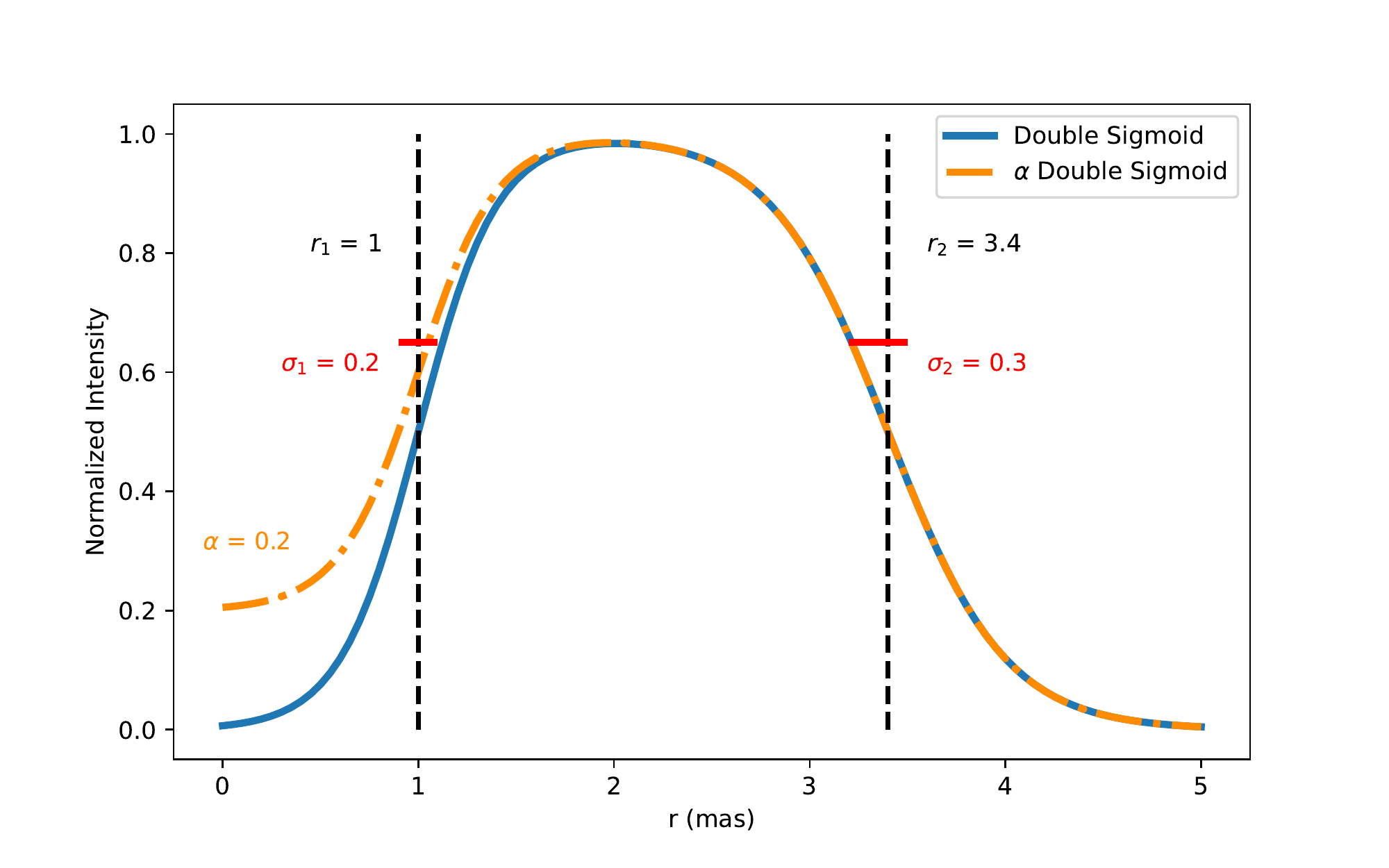}

\caption{We introduce a new function to describe the radial profile of the ring. The Double Sigmoid allows us to constrain the radii and their sharpness of the inner and outer rims separately. The blue curve corresponds to \autoref{eqn:DS} which, by definition, has no flux inside the inner rim. The orange dashed curve corresponds to the $\alpha$ Double Sigmoid from \autoref{eqn:aDS} which allows the center to have flux. The black dashed lines indicate the inner $r_1$ and outer $r_2$ radii of the profiles. The red horizontal lines indicate the $\sigma_1$ and $\sigma_2$ scale factors  }
\label{figure:sigmoid_example}
\end{figure}

We introduce two geometrical models with different profiles that we refer to as the Doughnut model and the Double Sigmoid model. The Doughnut model is a simple parabolic model that allows us to constrain the flux, inclination,  projection angle which indicates the direction of the major axis, spectral slopes, outer radius of the ring, and  thickness. The brightness profile has the following functional form:
\begin{equation}
    f(r) = 1-\left(\frac{r-\bar r}{2(r_{max}-r_{min})}\right)^2
\end{equation}

where $\bar r$, $r_{max}$, and $r_{min}$ are the mean, maximum and minimum radii, respectively.
The Double Sigmoid model adds more complexity to allow us to fit and scale the inner and outer radii independently, and the profile takes the following functional form:

\begin{equation}
  f(r) = \frac{1}{1+e^{\frac{-(r-R_{in})}{\sigma_{in}}}}* \frac{1}{1+e^{\frac{(r-R_{out})}{\sigma_{out}}}} 
\label{eqn:DS}
\end{equation}
where, $r_{in}$ and $r_{out}$ are the inner and outer radii, respectively, and $\sigma_{in}$ and $\sigma_{out}$ are their respective scale factors that set the sharpness of the edges. 
By definition, the Double Sigmoid model forces the inner cavity to be dark. To investigate the possibility of flux coming from the inner cavity, we modified the Double Sigmoid profile to take the following form:
\begin{equation}
  f(r) = \left( \alpha + \frac{1-\alpha}{1+e^{\frac{-(r-R_{in})}{\sigma_{in}}}}\right)* \frac{1}{1+e^{\frac{(r-R_{out})}{\sigma_{out}}}} 
\label{eqn:aDS}
\end{equation}

where the $\alpha$ term is allowed to vary to add flux to the center or go to zero if a completely dark center is the best fit. \autoref{figure:sigmoid_example} is an example of what that intensity profiles might look like using $R_{in}= 1$, $R_{out}=3.4$, $\sigma_{in}= 0.2$, $\sigma_{out}= 0.3$, and $\alpha$ = 0.2

\begin{figure}[H]
\centering
\includegraphics[trim={2cm 2cm 2cm 1
cm},clip,width=0.8\textwidth]{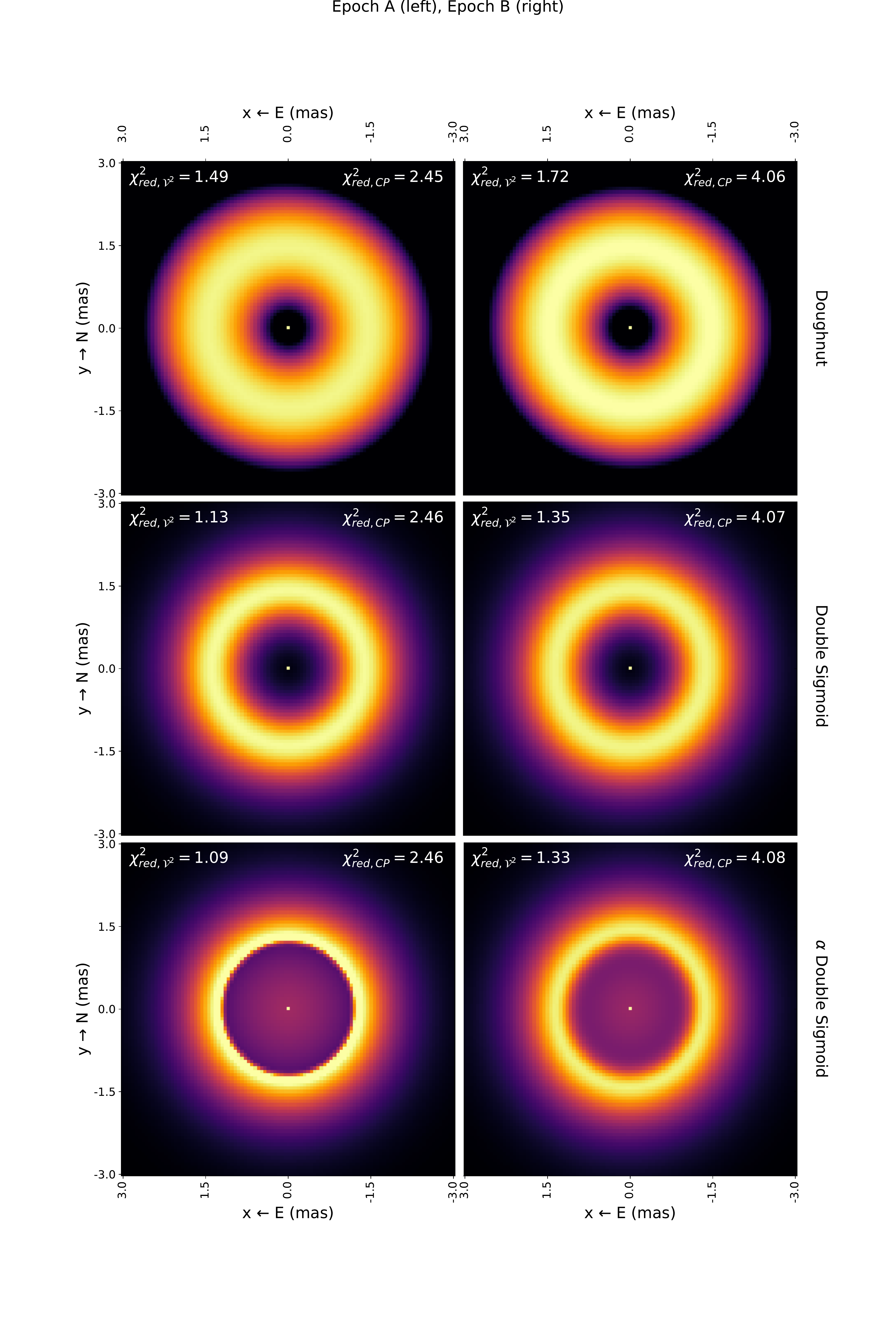}

\caption{Top row: Epoch A and B symmetric Doughnut models Epoch A $\chi^2_{red} = 1.49$ and Epoch B $\chi^2_{red} = 1.72$.\\ Middle row: Epoch A and B symmetric Double Sigmoid models $\chi^2_{red} = 1.13$ and Epoch B $\chi^2_{red} = 1.35$. Bottom row: Epoch A and B symmetric Double Sigmoid models $\chi^2_{red} = 1.09$ and Epoch B $\chi^2_{red} = 1.33$.}
\label{figure:donutsigmoidfig}
\end{figure}

\begin{figure}[H]
\centering
\includegraphics[trim={4.2cm 9cm 4cm 10cm},clip,width=0.9\textwidth]{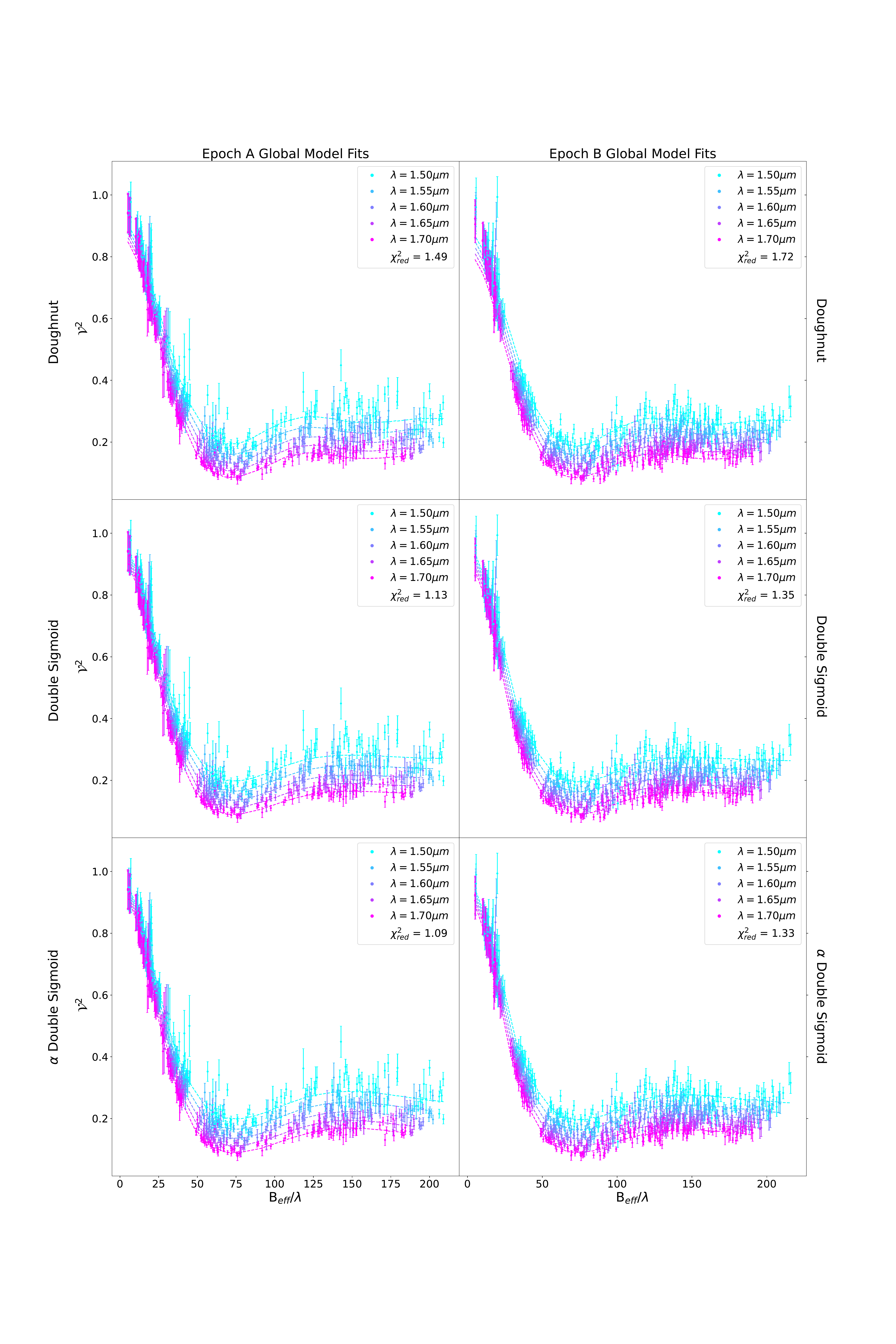}

\caption{Global fits of $\mathcal{V}^2$ for the Doughnut model (top row), the Double Sigmoid (middle row), and the $\alpha$ Double Sigmoid (bottom row). We see that the Doughnut model does the poorest job of fitting the short baselines while both Double Sigmoid models do a much better job. All three models fit the bounce at the longer baselines well which was expected since we see an inner cavity in all of them}
\label{fig:donutsigmoidfit}
\end{figure}

The full list of best-fit parameters from the three models can be seen compiled in \autoref{tab:bestfit}. Starting with the Doughnut model, as seen in the top two panels of \autoref{figure:donutsigmoidfig}, we notice that nearly all fitting parameters are consistent between epochs A and B. The disk is nearly face-on with a small inclination angle. The inner radii are very small, indicating a transition that is well beyond the expected dust wall. We found that around half of the flux originates from the disk, $~ 42.5\%$ comes from the star, and the rest is from an over-resolved halo component. We fixed the star's spectrum to be $\propto \lambda^{-4}$, and fit the spectra of the disk and halo as power laws. The spectral slope of the disk was fit to be $-0.10 \pm 0.14$, while the halo's slope was fit to be much steeper $2.08\pm1.16$. The goodness of these fits can be seen in Fig. 4, where the model is fitted to the squared visibility. The reduced (abbreviated `red') $\chi^2 $ for epoch A was calculated to be $\chi^2_{red,A} = 1.49$ and for epoch B $\chi^2_{red,B} = 1.72$, which is also indicated on the plots. Note that the $\chi^2_{red}$ values are based on $\mathcal{V}^2$ fits only. We see in the top two panels of \autoref{fig:donutsigmoidfit} that the model fits the longer baselines well, following the bounce and recovering an inner cavity like we expected as shown in the top two panels of \autoref{figure:donutsigmoidfig}. The shorter baselines, on the other hand, are not fit as well using this model. The Doughnut model relates the outer and inner radii and doesn't allow us to change the sharpness of the inner and outer rims. Seeing that the shorter baselines weren't fit very well using this model points to the larger scale features not being properly represented by the Doughnut. To get around this, we introduced the more elaborate Double Sigmoid model as discussed above.  

\begin{figure}[H]
\centering
\includegraphics[trim={0cm 0cm 2cm 1cm},clip,width=1\textwidth]{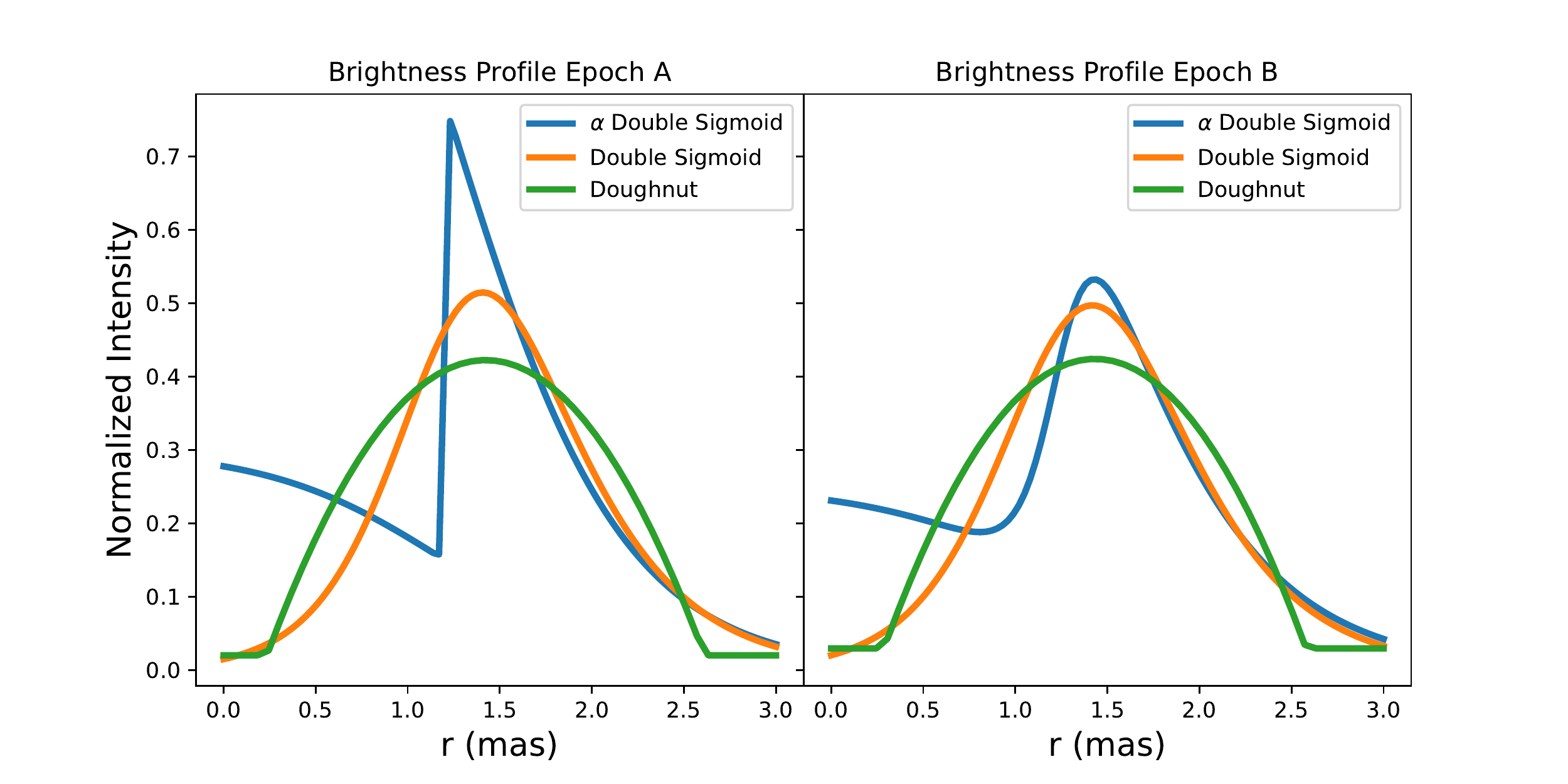}

\caption{Shown in green are the epoch A and B symmetric Doughnut brightness profiles normalized by the total flux. Both the inner and outer edges have some abruptness in the transition. We see a very narrow inner radius and a wide disk overall. Shown in orange are epochs A and B symmetric Double Sigmoid brightness profiles normalized by the total flux. Over-plotted in blue is the normalized profile of the $\alpha$ Double Sigmoid model. We see that the outer part of the profiles roughly matches while the inner parts differ. The $\alpha$ Double Sigmoid shows that 15$\%$ of the disk flux comes from the inner region and then shows a sharp inner edge transition}
\label{figure:donutsigmoidprofile}
\end{figure}

Similar to the Doughnut model, the Double Sigmoid model allows us to fit for the flux, inclination, and projection angle, spectral slopes, as well as inner and outer radii and their sharpness. The middle two panels of \autoref{figure:donutsigmoidfig} show the results of the symmetric Double Sigmoid model for epochs A and B. Similar to the Doughnut, the disk seems almost face-on, with a clear inner cavity. While the fitted inclination angle for A $i_A = 10.1^\circ\pm 2.9^\circ$ matches closely to the angle from the Doughnut fits, the inclination angle for epoch B $i_B = 18.1^\circ \pm 1.8^\circ$ is higher than expected. The inclination of the disk cannot change between epochs so we estimate the inclination to generally be $\lesssim 20^\circ$. The projection angles are significantly lower than the ones from the Doughnut model. That could be attributed to the almost face-on nature of the disk making it harder to constrain the projection angle. The inner rim being less diffuse and the outer rim being more diffuse in this model leads to a larger inner radius and smaller outer radius compared to the previous model. It is important to note that the $R_{in}$ and $R_{out}$ parameters do not necessarily reflect the size of the physical inner and outer radii. They are simply the best-fit parameters that reproduce the best brightness profile, and from those profiles, we estimate the physical radii based on the intensity transition. We show the normalized brightness profiles in \autoref{figure:donutsigmoidprofile} for both models and both epochs to get a better idea of how they compare. The Double Sigmoid profile is shown in orange while the Doughnut is shown in green. Indeed, we see the much smoother inner and outer rims of the Double Sigmoid compared to the sharp drops of the Doughnut model. The Double Sigmoid shows a slightly sharper inner edge compared to the outer edge. While the inner radius of the Double Sigmoid seems larger, and the outer seems smaller, due to the smoother edges, we can see from the profiles that the physical size of the disk from both models doesn't vary drastically, and the radii are not as different as the fitting parameters might lead us to believe. Other notable differences include the disk flux contribution increasing in the Double Sigmoid model, while the spectral slope decreases indicating an even smaller dependence on color. On the other hand, the halo flux contributions decreased, while the spectral slope grew redder, showing a stronger dependence on color.
 The reduced $\chi^2 $ for epoch A was calculated to be $\chi^2_{red,A} = 1.13$ and for epoch B $\chi^2_{red,B} = 1.35$, which is also indicated on the plots. We see in the middle two panels of \autoref{fig:donutsigmoidfit} that the model fits the longer baselines well, following the bounce and recovering an inner cavity like we expected as shown in the top middle panels of \autoref{figure:donutsigmoidfig}. Unlike the Doughnut model, the Double Sigmoid fits the shorter baselines much better. We clearly see that the $\chi^2_{red}$ significantly improves with the Double Sigmoid model where we see a bigger inner cavity and sharper inner rim. 
 
The $\alpha$ Double Sigmoid models can be seen in the bottom row of  \autoref{figure:donutsigmoidfig}. The initial global fit showed that adding the $\alpha$ term does improve the fit as evident by the $\chi_{red}^2$ decreasing for both epochs, $\chi_{red,A}^2=1.09$ and $\chi_{red,B}^2=1.33$ which can be seen on the bottom row of \autoref{fig:donutsigmoidfit}. This model tells us that $\sim 15\%$ of the light of the disk emerges from inside the rim. The normalized brightness profile can be seen in blue in \autoref{figure:donutsigmoidprofile}.

\begin{table}[H]
\centering
\caption{best fit results from $\mathcal{V}^2$ modeling}
\begin{tabular}{p{8mm}|p{6mm} p{8mm} p{9mm} p{10mm} p{10mm} p{9mm} p{8mm} p{6mm}p{6mm} p{6mm} p{8mm} p{9mm} p{5mm}p{5mm}}

    Model & 
    epoch & 
    incl. \newline (deg)&
    PA \newline (deg) &
    $R_{in}$ \newline (mas)  &
    $R_{out}$ \newline (mas)  &
    $\sigma_{in}$ \newline (mas)  &
    $\sigma_{out}$ \newline (mas)  &
    $\% f$\newline {star} & 
    $\% f$\newline{disk} & 
    $\% f$ \newline {halo} &
    disk \newline slope & 
    halo \newline slope & 
    \multicolumn{1}{l}{$\chi^{2}_{red}$}&
    $\alpha$\\

\hline

\multirow{3.5}{*}{\rotatebox[origin=c]{90}{Doughnut}} & \multirow{2}{*}{A} & 
        10.94 \newline $\pm 3.0$ & 
        42.30 \newline $\pm 10.0$ & 
        0.29 \newline $\pm 0.01$    & 
        2.60  \newline $\pm 0.02$    & 
         \hspace{2.5mm}-         & 
        \hspace{2.5mm}  -   & 
        43 \newline $\pm 0.5$    & 
        51 \newline $\pm 0.5$    & 
        6 \newline $\pm 0.5$     & 
        -0.10     \newline $\pm 0.1$    & 
        2.08     \newline $\pm 1.0$    & 
        \multirow{2}{*}{1.49} &      
        -
        \\ 
\cline{2-15}

 &\multirow{2}{*}{B} &
 
        10.05 \newline $\pm 3.0$ &  
        55.4  \newline $\pm 20.0$ &  
        0.35  \newline $\pm 0.01$ & 
        2.65  \newline $\pm 0.02$ & 
        \hspace{2.5mm}-            & 
        \hspace{2.5mm} -           & 
        42 \newline $\pm 0.5$   & 
        49 \newline $\pm 0.5$    &
        9 \newline $\pm 0.5$     & 
        -0.32  \newline $\pm 0.2$ & 
        2.10 \newline $\pm 1.0$   &  
        \multirow{2}{*}{1.72}&  
        -
        \\  

\hline
\multirow{5}{*}{\rotatebox[origin=r]{90}{\makecell{Double\\Sigmoid}}} & \multirow{2}{*}{A} &

        10.14 \newline $\pm 3.0$ &  
        3.4   \newline $\pm 10.0$ &  
        1.259 \newline $\pm 0.04$ & 
        1.252 \newline $\pm 0.05$ & 
        0.262 \newline $\pm 0.02$ & 
        0.426 \newline $\pm 0.02$ & 
        42  \newline $\pm 0.5$ & 
        54 \newline $\pm 0.5$ &
        4 \newline $\pm 0.5$  & 
        -0.07     \newline $\pm  0.1$ & 
        3.65    \newline $\pm 2.0$ &  
        \multirow{2}{*}{1.13}&  
        -
        \\  

\cline{2-15}

 &\multirow{2}{*}{B} &
 
        18.12 \newline $\pm 2.0$ &  
        9.58  \newline $\pm 6.0$ &  
        1.276 \newline $\pm 0.04$ & 
        1.268 \newline $\pm 0.05$ & 
        0.270 \newline $\pm 0.02$ & 
        0.452 \newline $\pm 0.02$ & 
        42 \newline $\pm 0.5$ & 
        54 \newline $\pm 0.5$ &
        4 \newline $\pm 0.5$  & 
        -0.25     \newline $\pm 0.2$ & 
        3.87     \newline $\pm 2.0$ &  
        \multirow{2}{*}{1.35}&  
        -
        \\  
        
\hline
\multirow{5}{*}{\rotatebox[origin=r]{90}{\makecell{$\alpha$ Double\\Sigmoid}}} & \multirow{2}{*}{A} &

        9.91 \newline $\pm 3.0$ &  
        0.2   \newline $\pm 20.0$ &  
        1.20 \newline $\pm 0.1$ & 
        1.20 \newline $\pm 0.2$ & 
        0.006 \newline $\pm 0.07$ & 
        0.479 \newline $\pm 0.02$ & 
        41 \newline $\pm 0.5$ & 
        56 \newline $\pm 0.5$ &
        3 \newline $\pm 0.5$  & 
        0.09     \newline $\pm  0.1$ & 
        4.56    \newline $\pm 3.0$ &  
        \multirow{2}{*}{1.09}&  
        $0.193$\newline $\pm 0.01$\\  

\cline{2-15}

 &\multirow{2}{*}{B} &
 
        19.07 \newline $\pm 2.0$ &  
        11.35  \newline $\pm 5.0$ &  
        1.251 \newline $\pm 0.02$ & 
        1.243 \newline $\pm 0.04$ & 
        0.103 \newline $\pm 0.04$ & 
        0.497 \newline $\pm 0.02$ & 
        41 \newline $\pm 0.5$ & 
        55 \newline $\pm 0.5$ &
        4 \newline $\pm 0.5$  & 
        -0.13     \newline $\pm 0.2$ & 
        4.62     \newline $\pm 3.0$ &  
        \multirow{2}{*}{1.33}&  
         $0.167$\newline $\pm 0.04$\\
  
\hline
\end{tabular}
\label{tab:bestfit}
\end{table}
\newpage

\newpage

\newpage

\begin{figure}[h]
\centering
\includegraphics[trim={1cm 1cm 3cm 1cm},clip,width=0.75\textwidth]{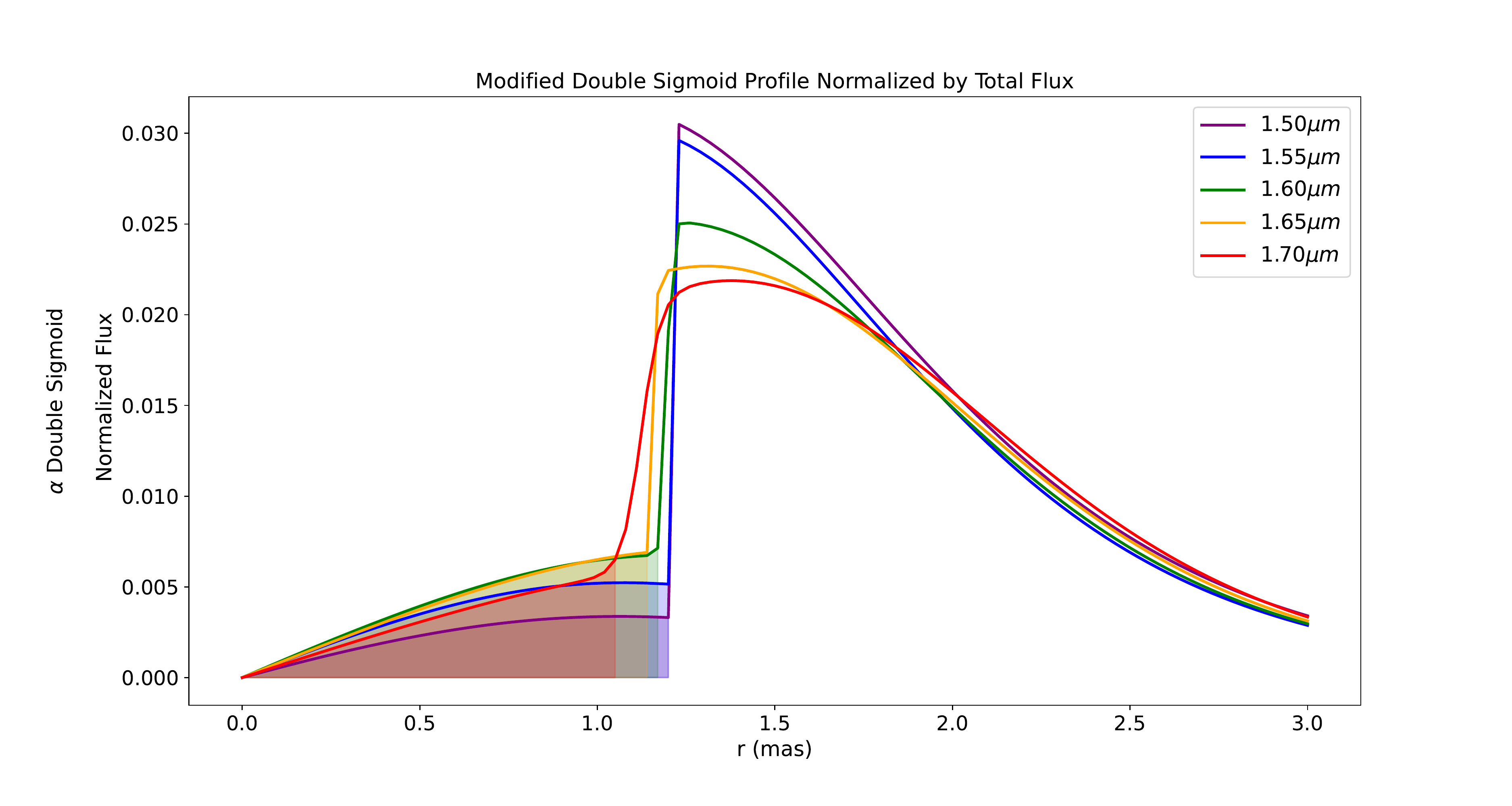}

\caption{Normalized flux profile of disk emission. The filled areas correspond to the fraction of flux contributed by inner emission we define as $\alpha$}
\label{figure:color}
\end{figure}

\begin{figure}[H]
\centering
\includegraphics[trim={0cm 0cm 1cm 
0cm},clip,width=0.65\textwidth]{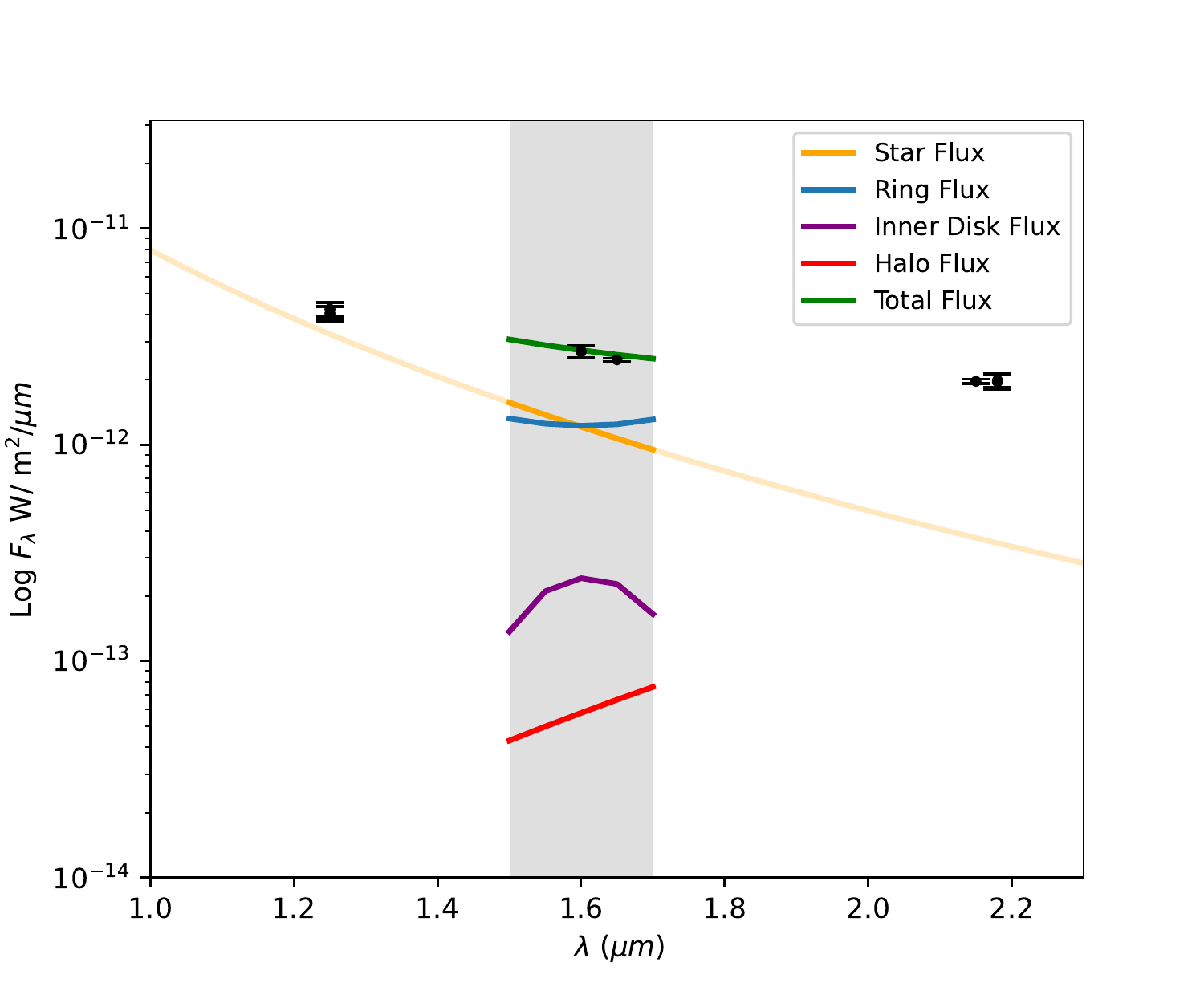}

\caption{Spectral energy distribution in J, H (shaded in gray), and K bands shown in the black dots. Over-plotted are the fitted spectra from epoch A. The star was assumed to have a $\lambda^{-4}$ spectrum which is shown in orange. The lighter orange line shows the extension of the slope across J and K bands. The blue line indicates the flux contribution from the ring, not including the inner emission. The inner emission flux is shown in purple. The over-resolved halo flux contribution is shown in red. The green line indicates the sum of all four contributions and it matches the observed photometry flux (SED data and references are in \autoref{table:photometry} in \autoref{appendix:sed}).}

\label{figure:sed}
\end{figure}

Next, to investigate the inner flux dependence on color, we fit each of the 5 wavelength channels independently using the same modified Double Sigmoid profile. We see in \autoref{figure:color} that all 5 wavelengths take a similar overall shape with varying flux contributions to the center. We normalized the profiles by the total flux and calculated the fractions of flux contributed by $\alpha$ to the total flux of the disk, which we show as the filled areas under the profile curves. These fractions allow us to calculate the spectrum of the inner disk independently of the spectrum of the outer disk. To calculate the inner disk spectrum, we used the inner flux fractions to calculate how much flux they contribute to the total light from the model that consists of the star, disk, and halo. We also calculated the spectra of the disk excluding the inner emission and the halo. We assumed the star's spectrum to be $\lambda^{-4}$. To test out the plausibility of the spectra, we added up the flux contributions from the star, disk, inner emission, and halo and compared the total to previously obtained photometry which we show in \autoref{figure:sed}. The spectral energy distribution (SED), which was built using SEDBYS \footnote{\url{https://gitlab.com/clairedavies/sedbys}} \citep{claire}, shows photometry data at J, H (shaded), and K band  with the spectra that we calculated for the four components of HD 190073 over-plotted. We see the slope of the total spectrum matches the data closely. The spectrum of the disk emission is rather flat which is more consistent with dust emission rather than the steeper free-free spectrum we would expect from gas emission. 
\newpage

\newpage
\begin{figure}[H]
\centering
\includegraphics[trim={0cm 3.2cm 0.5cm 
2.5cm},clip,width=0.72\textwidth]{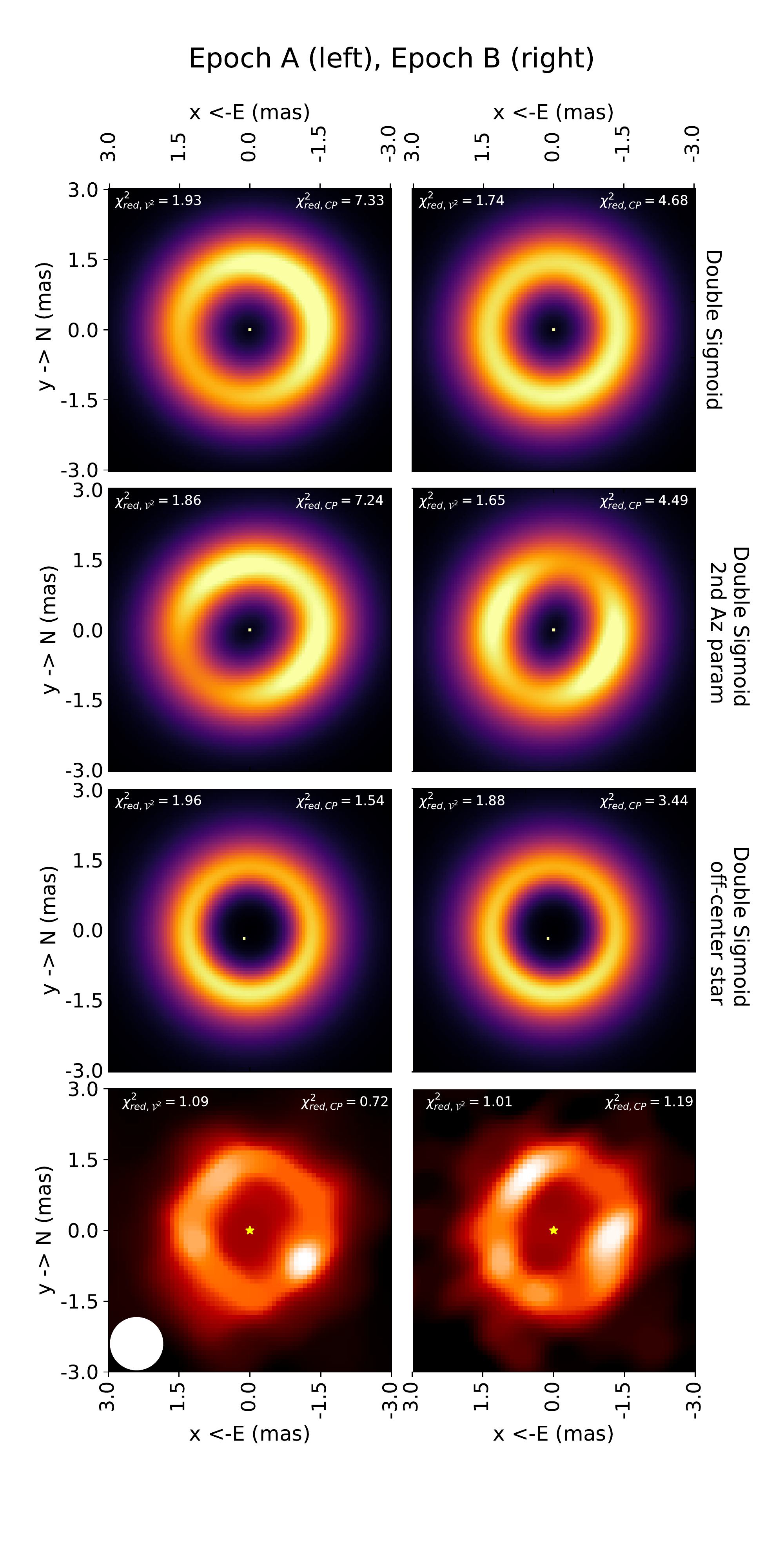}

\caption{Top row: Epoch A and B Skewed Double Sigmoid ring model. Second row: Epoch A and B higher order Skewed Double Sigmoid ring model. Third row: Epoch A and B off-center star Double Sigmoid model. Bottom row: Epoch A and B model-free images. The white circle in the left bottom corner of the left panel corresponds to the effective beam size $\lambda/(2B_{max}) = 0.55$ mas where $\lambda =1.65$  $\mu$m and $B_{max}$ = 300 m }
\label{figure:asymmetric}
\end{figure}
\newpage
\subsection{Asymmetric Modeling}

So far, we have been assuming that the disk is symmetric and therefore only fitting the squared visibilities. However, as mentioned before, the closure phases are non-zero and non-$180^\circ$ which means that there is asymmetry in the disk. In this section, we use 3 different simple geometric models to attempt to characterize the asymmetry.

These simple models are no longer valid in describing the geometry of the disk so we have to introduce asymmetry parameters. Azimuthal asymmetry is expected from flared disks seen at an inclination. Since the Double Sigmoid performed better than the Doughnut in the symmetrical case, we used a Double Sigmoid skewed ring model with azimuthal asymmetry \citep{Monnier2006} as the base of our modeling and added different asymmetries to it. We tested multiple azimuthal asymmetries such as adding a point source free parameter, adding a higher order multipole for azimuthal asymmetry, considering an off-center star, and more.

These skewed models allow us to vary the brightness distribution across the disk. In this paper, we present three of the models we tested and show how their closure phase $\chi^2_{red,CP}$ compare to give us a better picture of the disk. A full list of the best-fit parameters from these three models can be found compiled in \autoref{table:asym} in \autoref{appendix:asym}. Starting with a simple skewed Double Sigmoid ring, we see in the top two panels of \autoref{figure:asymmetric} that one side of the disk tends to be brighter. This model introduces two additional fitting parameters to represent the harmonic azimuthal variation. The amplitude is referred to as ``Az Amp$_{1}$" in \autoref{table:asym} in \autoref{appendix:asym}, and the projection angle, which is defined with respect to the global projection angle is denoted as ``Az PA$_1$". Both parameters have a subscript that represents their order. The amplitude of variation is stronger in epoch A compared to B. The inclination of the flared disk causes the side farther to us to appear brighter but this skewness does not produce a good fit as indicated by $\chi^2_{red,CP,A} = 7.33$ and $\chi^2_{red,CP,B} = 4.68$.  This model is not complex enough to model the asymmetries, so in the next model, we introduced a higher order of asymmetry, a $\sin(2\theta)$, to the ring. The higher order allows us to model simple structures that might be in the disk. This adds an extra pair of fitting parameters, Az Amp$_{2}$ and Az PA$_2$, which we list in \autoref{table:asym} as well. Models of the disk, which can be seen in the second row of \autoref{figure:asymmetric}, produce two distinct bright regions, and the CP  $\chi^2_{red,CP}$ has decreased but not by much.  $\chi^2_{red,CP,A} = 7.24$ and $\chi^2_{red,CP,B} = 4.49$ for epochs A and B respectively. The first-order azimuthal amplitudes stay consistent with the previous model. Interestingly enough, the second-order amplitudes have the opposite strengths for the two epochs. Epoch A Az Amp$_{A,2}$ is lower than Az Amp$_{A,1}$ and epoch B Az Amp$_{B,2}$ is higher than Az Amp$_{B,1}$.  We are not giving significant physical meanings to these models because they are meant to describe a complex structure with simple parameters. Instead of introducing even higher orders, we went with a less complex approach by testing a skewed Double Sigmoid with an off-center star, which can be seen in the third row of \autoref{figure:asymmetric}. This model favored a slightly off-center star and dropped the CP $\chi^2_{red,CP}$ significantly. The simple skewed ring started with a $\chi^2_{red,CP,A} = 7.33$ and $\chi^2_{red,CP,B} = 4.68 $, which dropped slightly by adding the second azimuthal parameter to be $\chi^2_{red,CP,A} = 7.24$ and $\chi^2_{red,CP,B} = 4.49 $, and then dropped sharply with the addition of the off-center star to be $\chi^2_{red,CP,A} = 1.54$ and $\chi^2_{red,CP,B} = 3.44 $. The amplitude of skewness of the two epochs almost matches in this model. Two extra parameters were introduced to characterize how much the star was fit away from the center, an x offset and a y offset. The x offsets are similar for both epochs and they place the star $~0.1 \pm 0.01$ mas west of the center. The y offsets both placed the star $~0.1 \pm 0.01$ mas south of the center. While none of the asymmetric models fit to $\chi^2_{red}\sim1$, we are inclined to favor the off-center star as being at least a plausible component. The reason for the star's off-center location has not been established yet. One of our leading theories is the presence of a binary in the center, which will need further studies.

\section{imaging}
\label{sec:imaging}

Since we were not getting a good fit using simple models, we decided to try imaging in the hopes of unlocking some of the mystery. As mentioned in \autoref{sec:Modeling}, the closure phase data show non-zero and non-180$^\circ$ measurements that indicate some form of asymmetry in the disk. However, our asymmetric models did not succeed in characterizing the asymmetry. One possibility is an off-center star, but the physical reason behind that is still not established. Since the goal of imaging is to search for finer-scale details, we wanted to fine-tune the data binning in the two epochs. As mentioned in \autoref{subsec:combine}, CHARA can resolve the motion of one resolution element every 13 days. However, the first night in epoch A is more than 13 days away from the other two nights. Similarly, the last night in epoch B is more than a month away from the first two nights. Therefore, we removed the 2019-05-09 night from epoch A and 2019-08-27 from epoch B. 

Images were reconstructed using conventional regularized maximum likelihood and the open-source OITOOLS.jl
package 
\footnote{\url{https://github.com/fabienbaron/OITOOLS.jl}}.
Only powerspectra and closure phases were used for the reconstructions. In line with previous mixed modeling and imaging works \citep[e.g. SPARCO in][]{Kluska2014}, the star was modeled as a point source in the center of the field of view, of spectrum proportional to $\lambda^{-4}$, while the environment was assumed to follow a $\lambda^\alpha$ dependency, with $\alpha$ free to vary. In addition, we modeled potential extended or background emission as a zero visibility component, also following $\lambda^{-4}$. Thus there were three free parameters in addition to image pixels: $\alpha$ and the flux ratios disk/star and extended emission/star.

Minimization was ensured by the gradient-based VMLM-B \citep{Thiebaut2002} algorithm from OptimPack, enforcing positivity of the disk intensity. Three other priors were employed in addition to positivity: compactness to regularize large-scale
emission, total variation to handle small-scale pixel correlations \citep[see e.g.][for the mathematical expression of these]{Thiebaut2017}, as well as a novel azimuthal regularization to ``circularize" YSO disks at medium scales. Given four hyperparameters (the centroid, inclination and position angle of the disk), this regularizer simply computes the location of concentric elliptical rings and then sums the azimuthal pixel variances along them. Minimizing the regularizer enforces relative smoothness along the rings, but unlike total variation. Since the expression can be written as a squared $\ell_2$ norm of a linear operator on the image, it lends itself well to the gradient-based approach. Hyperpriors for the four hyperparameters were uniform and an initial hyperparameter range was roughly determined
from the data by directly minimizing the variance of the de-rotated
powerspectra. The Nelder-Mead simplex \citep{nelder_simplex_1965} was used to optimize the four
hyperparameters, new reconstructions were run with different sets of
hyperparameters until a global minimum was found. Values for the other
regularization hyperparameters were chosen based on quick simulations
(copying the (u,v) coverage and signal-to-noise of the actual data and
reconstructing disks), within the expected range \citep{Renard2011}.

\begin{figure}[H]
\centering
\includegraphics[width=\textwidth]{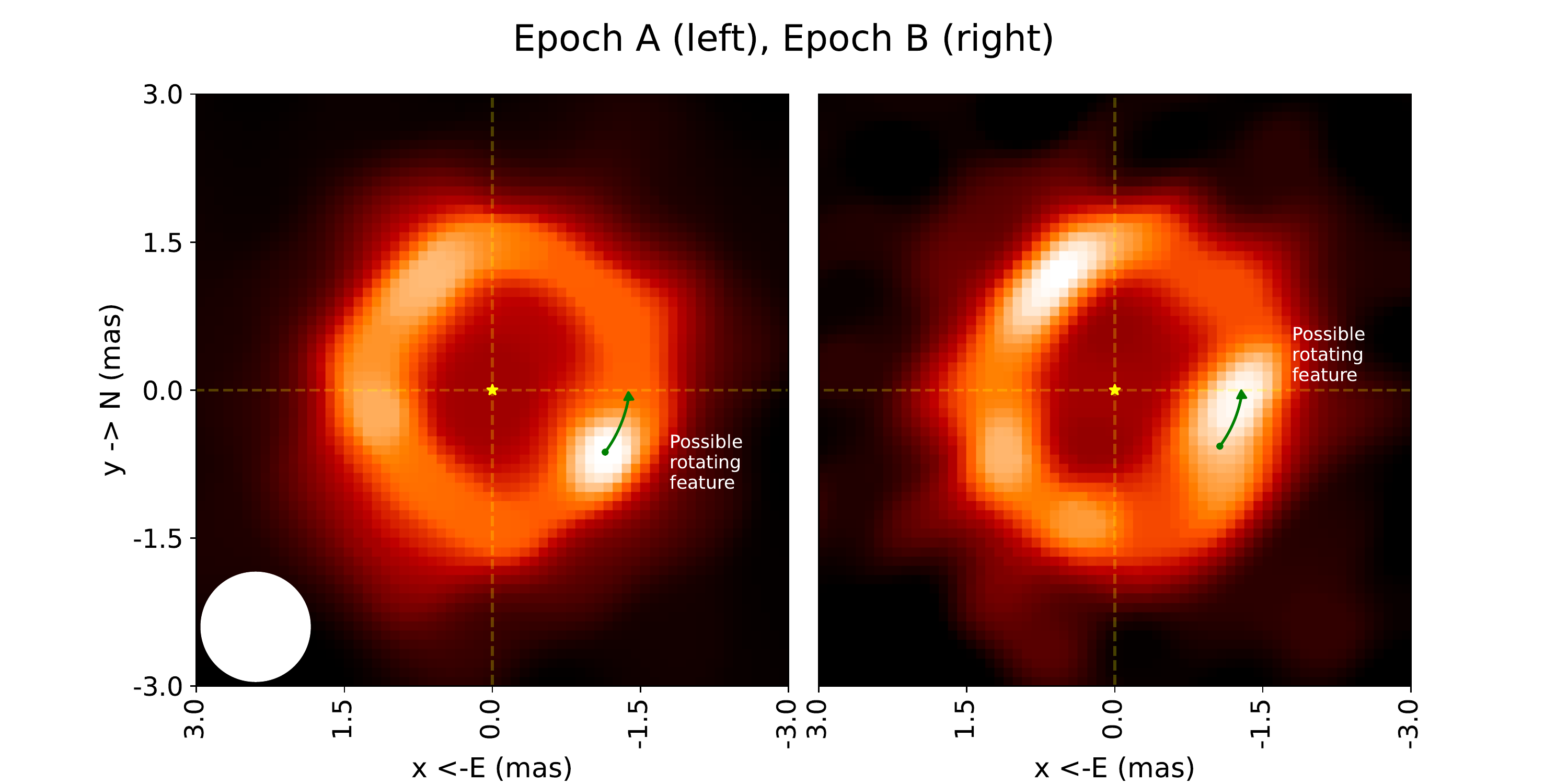}

\caption{Epochs A and B images. The dashed crosshairs show the center of the image. While the star is in the center, the disk is not which shows that this matches our off-center star model. The green arrow indicates a 27$^\circ$ rotation. The spot starts at the bottom of the green arrow in epoch A and moves to the top in epoch B. The white circle in the left bottom corner of the left panel corresponds to the effective beam size $\lambda/(2B_{max}) = 0.55$ mas where $\lambda =1.65$  $\mu$m and $B_{max}$ = 300 m }
\label{fig:images}
\end{figure}

 We see a face-on thin ring with some inner emission, similar to the $\alpha$ Double Sigmoid model, and an off-center star. We clearly recover a cavity in the center where we see some fine-scale structure that seems to change. The last row of \autoref{figure:asymmetric} shows the images that we produced without priors from modeling. The bright spots in the ring seem to correspond with the location of the bright areas seen in a combination of the skewed models. Note that in the images, the point source representing the star is shown in the center of the image while the disk's center is off-center. The final reduced $\chi^2$ for the images is $\chi^2_\text{v2} \simeq
1.09$ and $\chi^2_\text{t3phi}\simeq 0.72$ for epoch A and $\chi^2_\text{v2} \simeq
1.01$ and $\chi^2_\text{t3phi}\simeq 1.19$ for epoch B. A closer look at the images can be seen in \autoref{fig:images}. We explored if the changes might be due to rotation. We performed a qualitative analysis of the apparent rotation between epochs A and B by manually rotating the disk to match the bright feature in the bottom right quadrant in epoch A to the bright spot on the middle right edge of the ring in epoch B which is indicated in  \autoref{fig:images}. The arrows represent a $27^\circ$ counterclockwise rotation on the disks. Through modeling we found that the $\alpha$ Double Sigmoid intensity peaked at a radius $R_{A} =1.23\pm0.02$ mas and  $R_{B} =1.5\pm0.02 $ mas. The Keplerian rotational period at this smaller radius that we average to be $R = 1.35$ mas is 182.7 days. The two epochs are 32 days apart which corresponds to a 63$^\circ$ rotation based on the calculated period. However, this motion is not consistent with the 27$^\circ$ rotation we estimate from the images which is more than twice as slow as Keplerian. The origin of such features is not well-known but we speculate that the motion could be caused by interactions in the outer disk versus an object embedded in the inner disk, for example. This structure could be related to planet formation, instabilities in the accretion flow, or even magnetic fields. This change could also be an artifact of the (u,v) coverage and not a physical process at all so additional epochs are needed to confirm the origin of these changes.

\section{Discussion}
\label{sec:disc}
The symmetric $\mathcal{V}^2$-only models revealed a clear inner cavity in the disk, along with rough radii estimates. We are taking the radius at which the intensity peaks in the Double Sigmoid and $\alpha$ Double Sigmoid models to be representative of the evaporation front. Since the inner radius is set by the dust evaporation temperature, we can use the average radius from the models to can estimate the temperature of the dust wall using the following equation from \citet{Dullemond2010}.

\begin{equation}
T_{\text {dust }}=T_{*} \frac{1}{\epsilon^{1 / 4}} \sqrt{\frac{R_{*}}{2 R}}
\end{equation}
This equation treats the inner rim as an optically thick wall that radiates like a blackbody, assumes that the gas inside the rim is transparent, and includes the “backwarming” of dust grains. The $\epsilon$ term is the ratio of the effectiveness of emission at the wavelength
at which the dust radiates away its heat and absorption at stellar wavelengths. Assuming large dust grains, we can set $\epsilon = 1$ and solve for the temperature of the inner edge using values from 
\autoref{table:stellar params} and $R = 1.35\pm0.2$ mas for the radius of the inner rim. We calculate an inner rim temperature of T$_{dust} = 1367.13 \pm 107$ K which is a reasonable estimate. If the dust grains have a larger $\epsilon$, meaning that they cool more efficiently, then the grains can exist closer in to the star. Our modified $\alpha$ Double Sigmoid model showed that $\sim 15\%$ of the disk flux comes from the inside of the ring. The sub-AU inner emission could be due to such grains with $\epsilon >1$, gas emission, or grains that can survive at higher temperatures.

 We have non-zero closure phases which indicate that we have some asymmetry in the ring. The closure phases are relatively small so the asymmetry is not huge. From our fits, it seems that the asymmetry is best fit by including an off-center star. While not certain about the physical meaning of that fit, one speculation is that we are looking at a binary star system. 
 
 One reassuring confirmation that the model is probable, is that the imaging produced an off-center star as well. The imaging also confirmed the cavity in the center of the ring and showed new evidence of small-scale structures with rotation slightly slower than Keplerian at that radius. There is already evidence of this temporal variation on the inner few AU scale in protoplanetary disks. \citet{kobus2020} showed evidence of interferometric temporal variation occurring in $>10\%$ of their sample of 68 accretion disks. The asymmetric variations could be attributed to companions, forming exoplanets, asymmetries in the dust density distribution,  asymmetric illuminations of the circumstellar material as a result of stellar spots or obscuring, or artifacts. There are a few inner disk dynamics that have been observed that could help unlock the mystery of the variations we see in our disk. Some possible inner disk dynamics include self-shadowing by the inner disk on the outer disk \citep{Garufi2022}, dippers caused by turbulent accretion \citep{Alencar2010}, or even a small planet actively accreting close to the inner rim with accompanying spirals, vortices, or misalignment (see \citet{Benisty2017}, \citet{Marr2022}). Characterizing these structures is going to be essential to furthering our understanding of planet formation.

\section{Conclusions}
We tested multiple simple geometric models, both symmetric and asymmetric, in hopes of getting a better glance at the young Herbig Be star HD 190073's accretion disk. With more sensitive data, better (u,v) coverage, and a shorter time span to avoid smearing effects, we were able to model and image the disk with unprecedented accuracy. CHARA's long baselines probe the finer structures of the disk, while the short baselines of VLTI look at the larger-scale structures. For a rapidly changing object like HD 190073, it is important to get near-simultaneous observations with CHARA and VLTI to get as close to an instant snapshot of the disk as possible. While there is still some mystery surrounding HD 190073, we were able to produce convincing evidence of a ring-like structure with an inner cavity, and some evidence of changes. The modified $\alpha$ Double Sigmoid model showed that the inner and outer rims are relatively smooth, with the inner rim having a sharper profile due to the dust destruction front. We also see that the disk is almost face-on with inclination angles $i \lesssim 20^\circ$. The modeling revealed the fractions of flux contributed by each component. The star contributes $41\%$, the halo $4.6\%$, and the disk $\sim 56\%$ with $\sim15\%$ of that disk flux coming from inside the inner rim. We found that the best simple model to fit the closure phase measurements is a skewed Double Sigmoid ring with an off-center star, which was backed up by the imaging. The imaging revealed possible evidence of rotating sub-AU features. They seem to rotate $27^{\circ}$ counterclockwise in the span of the 32 days between epochs A and B which would correspond to a rotational period that is more than two times slower than Keplerian. This change could be caused by dynamics in the outer disk, or it could be an artifact.

We plan on continuing the study of HD 190073 in the future. We have new MYSTIC \citep{Monnier2018} K-band data that we will be adding to get a wider look at the disk. We also plan on quantitatively analyzing the temporal rotation to get a better estimate of the rotation angle and to measure the correlation between the two epochs. The correlation will be able to further confirm that the rotation is not an artifact of the (u,v) coverage. Future data will also reveal if the star is off-center due to an inner binary.


\label{sec:conclusions}


\acknowledgments
This work is based upon observations obtained with the Georgia State University Center for High Angular Resolution Astronomy Array at Mount Wilson Observatory.  The CHARA Array is supported by the National Science Foundation under Grant No. AST-1636624 and AST-2034336.  Institutional support has been provided by the GSU College of Arts and Sciences and the GSU Office of the Vice President for Research and Economic Development. MIRC-X received funding from the European Research Council (ERC) under the European Union's Horizon 2020 research and innovation programme (Grant No. 639889). JDM acknowledges funding for the development of MIRC-X (NASA-XRP NNX16AD43G, NSF-AST 1909165) and MYSTIC (NSF-ATI 1506540, NSF-AST 1909165). This research has made use of the Jean-Marie Mariotti Center Aspro and SearchCal services. S.K., N.A., and C.L.D.  acknowledge support from an ERC Starting Grant (``ImagePlanetFormDiscs'', grant agreement No.\ 639889), ERC Consolidator Grant (``GAIA-BIFROST'', grant agreement No.\ 101003096), and STFC Consolidated Grant (ST/V000721/1). A.L. received funding from STFC studentship No.\ 630008203. Observing travel support was provided by STFC PATT grant ST/S005293/1. Based on observations collected at the European Organisation for Astronomical Research in the Southern Hemisphere under ESO programmes 0101.C-0896(A) and 0103.C-0915(A,B,C).  E.A.R. and J.D.M. acknowledges
support from NSF AST 1830728.

%
%



\vspace{5mm}
\facilities{CHARA,VLTI}
\software{Astropy \citep{astropy:2013, astropy:2018, astropy:2022}, PMOIRED \citep{2022merand}, SEDBYS \citep{claire}, \textsc{PNDRS} \cite[v3.52;][]{LeBouquin2011}, OITOOLS.jl 
(\url{https://github.com/fabienbaron/OITOOLS.jl}) }


\bibliographystyle{aasjournal}
\bibliography{references}
\newpage
\begin{appendices}

\section{SED data table }
\label{appendix:sed}
\begin{table}[H]
\centering

\caption{HD 190073 photometry data}
\begin{tabular}{llll}

Wavelength & $\lambda F_{\lambda}$   & Date & Reference\\
$\mu$m     & $10^{-13}$\,W\,m$^{-2}$ &      &   \\
\hline

1.25 & $53.036\pm3.908$ & 2006Jun02 & \citet{Tannirkulam2008me} \\ 
1.25 & $48.370$ & 2000Jun02 & \citet{Lazareff2017} \\ 
1.25 & $50.676\pm0.887$ & 2000Jun04 & \citet{Cutri2003fp} \\ 
1.60 & $42.956\pm2.769$ & 2006Jun02 & \citet{Tannirkulam2008me} \\ 
1.60 & $43.353$ & 2000Jun02 & \citet{Lazareff2017} \\ 
1.65 & $40.817\pm0.639$ & 2000Jun04 & \citet{Cutri2003fp} \\ 
2.15 & $42.304\pm1.052$ & 2000Jun04 & \citet{Cutri2003fp} \\ 
2.18 & $42.629$ & 2000Jun02 & \citet{Lazareff2017} \\ 
2.18 & $43.421\pm3.199$ & 2006Jun02 & \citet{Tannirkulam2008me} \\ 

\hline

\end{tabular}
\label{table:photometry}

\end{table}
\section{Asymmetric Modeling Fitting Parameters }
\label{appendix:asym}

\begin{table}[!ht]
\caption{Fitting parameters results from asymmetric models}
\label{table:asym}

    \centering
    \begin{tabular}{c|c|c|c|c|c|c}
   
        Model & \multicolumn{2}{c|}{Skewed Ring} & \multicolumn{2}{c|}{2nd Az Parameter} & \multicolumn{2}{c}{Off-Center} \\ \hline
        Epoch & A & B & A & B & A & B \\ \cline{2-7}
         incl. (deg) & 14.18 ± 3.0 & 16.4 ± 3.0 & 25.2 ± 2.0 & 30.0 ± 2.0 & 10.7 ± 3.0 & 19.7± 2.0 \\ 
        PA(deg)  & 101.7 ± 10.0 & 162.1 ± 10.0 & 118.2 ± 5.0 & 157.9 ± 5.0 & 3 ± 10.0 & 157.7 ± 6.0 \\ 
        R$_{in}$(mas)  & 1.38 ± 0.07 & 1.27 ± 0.06 & 1.39 ± 0.07 & 1.26 ± 0.05 & 1.16 ± 0.02 & 1.11 ± 0.02 \\ 
        R$_{out}$ (mas)  & 1.37 ± 0.08 & 1.26 ± 0.06 & 1.39 ± 0.07 & 1.25 ± 0.06 & 1.14 ± 0.04 &  1.10 ± 0.04 \\ 
        $\sigma_{in}$ (mas)  & 0.30 ± 0.03 & 0.28 ± 0.03 & 0.31 ± 0.03 & 0.28 ± 0.03 & 0.18 ± 0.01 & 0.17 ± 0.02 \\ 
        $\sigma_{out}$ (mas) &  0.40 ± 0.02 & 0.46 ± 0.03 & 0.43 ± 0.03 & 0.52 ± 0.03 & 0.44 ± 0.01 & 0.53 ± 0.02 \\ 
        \% f$_{star}$ & 42 ± 0.5 & 42 ± 0.5 & 42 ± 0.5 & 42 ± 0.5 & 42 ± 0.5 & 42 ± 0.5 \\ 
        \% f$_{disk}$ & 54 ± 0.5 & 54 ± 0.5 & 54 ± 0.5 & 55 ± 0.5 & 54 ± 0.5 & 54 ± 0.5 \\ 
        \%f$_{halo}$ & 4 ± 0.5 & 4 ± 0.5 &  4 ± 0.5 & 3 ± 0.5 & 4 ± 0.5 & 4 ± 0.5 \\ 
        disk slope  & -0.14 ± 0.2 & -0.27 ± 0.2 & -0.12 ± 0.2 & -0.22 ± 0.2 &  -0.13 ± 0.1 & -0.20 ± 0.2 \\ 
        halo slope  & 3.60 ± 3.0 & 4.1 ± 3.0 & 3.58 ± 3.0 & 4.48 ± 4.0 & 4.10 ± 2.0 & 3.4 ± 3.0 \\ 
        Az Amp$_1$  &  0.14 ± 0.01 & 0.04 ± 0.01 & 0.14 ± 0.01 & 0.05 ± 0.01 & 0.08 ± 0.02 & 0.07 ± 0.01 \\ 
        Az PA$_1$ (deg)  & -153.2 ± 10.0 & 57.1 ± 20.0 & -170.9 ± 6.0 & 38.6 ± 10.0 & -170.4 ± 20.0 & 39.57 ± 9.0 \\ 
        Az Amp$_2$ & - & - & 0.09 ± 0.02 & 0.15 ± 0.02 & - & - \\ 
        Az PA$_2$ (deg)  & - & - & 80.05 ± 4.0 & -92.4 ± 3.0 & - & - \\ 
        x offset (mas) & - & - & - & - & 0.12 ± 0.01 & 0.09 ± 0.01 \\ 
        y offset (mas) & - & - & - & - & -0.15 ± 0.01 & -0.09 ± 0.01 \\ 
        Overall $\chi^2_{red}$ & 2.10 & 2.91 & 2.08 & 2.83 & 1.35 & 2.28 \\
        $\mathcal{V}^2$ $\chi^2_{red}$ & 1.93 & 1.74 & 1.86 & 1.65 & 1.96 & 1.88 \\
        CP $\chi^2_{red}$ & 7.33 & 4.68 & 7.24 & 4.49 & 1.54 & 3.44 \\\hline
    \end{tabular}

\end{table}
\end{appendices}


\listofchanges

\end{document}